\documentclass{eas}
\usepackage{graphicx}
\newcommand{\spitzer}{{\it Spitzer}}

\newcommand{\micron}{\mbox{$\mu$m}}

\newcommand{\farcs}{\mbox{$^{\prime\prime}$.}}

\begin{document}

%%-----------------------------
%%      the top matter
%%-----------------------------
\title{Revolutionizing our View of Protostellar Multiplicity and Disks: 
The VLA Nascent Disk and Multiplicity (VANDAM) Survey of the Perseus Molecular Cloud
} 
\runningtitle{VANDAM Survey}
\author{John J. Tobin}\address{Leiden Observatory}
\author{Leslie W. Looney}\address{University of Illinois}
\author{Zhi-Yun Li}\address{University of Virginia}
\author{Claire J. Chandler}\address{National Radio Astronomy Observatory}
\author{Michael M. Dunham}\address{Harvard-Smithsonian Center for Astrophysics}
\author{Dominique Segura-Cox$^2$}
\author{Erin G. Cox$^2$}
\author{Robert J. Harris$^2$}
\author{Carl Melis}\address{University California San Diego}
\author{Sarah I. Sadavoy}\address{Max Planck Institute for Astronomy}
\author{Laura P\'erez$^4$}
\author{Kaitlin Kratter}\address{University of Arizona}
\begin{abstract}
There is substantial evidence for disk formation taking place during the
early stages of star formation and for most stars being born in multiple systems; however,
protostellar multiplicity and disk searches have been hampered by low resolution, sample
bias, and variable sensitivity. We have conducted an unbiased, high-sensitivity
Karl G. Jansky Very Large Array (VLA) survey toward all known protostars (n = 94) in the Perseus 
molecular cloud (d $\sim$ 230 pc), with a resolution of $\sim$15 AU (0.06$^{\prime\prime}$) at
$\lambda$ = 8 mm. We have detected candidate protostellar disks toward 
17 sources (with 12 of those in the Class 0 stage) and we have 
found substructure on $<$ 50AU scales for three Class 0 disk candidates, 
possibly evidence for disk fragmentation. We have discovered 16 new 
multiple systems (or new components) in this survey; 
the new systems have separations $<$~500~AU and 3 by $<$~30~AU. We also 
found a bi-modal distribution of separations, with peaks at $\sim$75 AU and $\sim$3000 AU,
suggestive of formation through two distinct mechanisms: disk and turbulent fragmentation.
The results from this survey 
demonstrate the necessity and utility of uniform, unbiased surveys of 
protostellar systems at millimeter and centimeter wavelengths.

\end{abstract}
\maketitle
%%-----------------------------
%%      your text
%%-----------------------------
\section{Introduction}
Stars form due to the gravitational collapse
of dense cores within molecular clouds. Conservation
of angular momentum in this infalling material causes the formation
of a rotationally-supported disk around the nascent protostar. However, this picture
may be complicated by the presence of magnetic fields that can remove angular
momentum from the infalling material (Allen et al. 2003). In a similar
vein, wide multiple protostellar systems can form via rotational breakup of the
collapsing cloud (e.g., Burkert and Bodenheimer 1993). Turbulent fragmentation (Padoan et al. 2007)
has recently become a favorable route for the formation of both wide and close multiples;
the close multiples migrate inward from initially larger separations (e.g., Offner et al. 2010). 
Close multiples may also form by fragmentation of 
a massive disk via gravitational instability (e.g., Adams et al. 1989); however, 
it is unknown if disks of sufficient mass and radius form in young protostellar systems.

Thus far, sub/millimeter studies of Class 0 protostars have not had the resolution
to resolve the scale of disks and close multiples, and most samples have been small and/or biased.
To make a substantial leap in our knowledge of both protostellar disks and multiplicity, we have conducted the VLA
Nascent Disk and Multiplicity (VANDAM) Survey toward all known protostars in the Perseus molecular
cloud. The survey was conducted in A and B configurations with the VLA
at 8 mm, 1 cm, 4 cm, and 6.4 cm, observed only in wide-band continuum and 
reaching a high spatial resolution of  0\farcs065 (15 AU) at 8 mm. 
We will focus only on the 8 mm and 1 cm results in this contribution.
Our sample is drawn primarily from the \spitzer\ survey by Enoch et al. (2009) as well as
all known candidate first hydrostatic core objects and other deeply embedded sources.

\section{Protostellar Disks}
We have detected resolved structures that are consistent with protostellar disks toward 12/43 Class 0 sources 
and 5/37 Class I sources. Examples of a few Class 0 disk candidates are shown in Figure 1. Some of the
other candidates are further analyzed in Segura-Cox et al. (submitted) and the apparent circumbinary disk candidates
are shown in Tobin et al. (submitted). Of the candidates shown, they appear to have 
radii of 30 AU or less. To determine if we are tracing the full extent of the disks at 8 mm, we 
examined both ALMA 870 \micron\ data and VLA 7 mm data toward the known Class 0 disk 
around L1527 IRS (Tobin et al. 2012), shown in Figure 2. The 7 mm data from the VLA have a resolution 
of $\sim$0\farcs25, but are convolved with the same restoring beam as the ALMA data; 
the 7 mm data recover all the flux observed in the most compact VLA configuration
(Melis et al. 2011). The deconvolved sizes are 0\farcs62 and 0\farcs26 at 870 \micron\ and 7 mm, respectively, 
demonstrating that this disk has a smaller apparent radius at 7 mm. We have further
verified the more compact 7 mm emission using visibility amplitude profiles. Thus, 8 mm disk
sizes may be lower limits.

The smaller extents at longer wavelengths have been seen toward more-evolved Class II objects (P\'erez et al. 2012)
and this is interpreted as the inward radial drift of the dust grains due
to aerodynamic drag (Weidenschilling 1977). Models of dust growth and migration by Birnstiel et al. (2010) show that 
this effect may happen quickly enough to be apparent while still in the protostellar phase. We cannot, however, rule-out
grain growth alone in the inner disk being responsible for this effect. Multi-wavelength modeling will be necessary to 
assess the difference in dust emission from short to long wavelengths.

\begin{figure}
\begin{center}
\includegraphics[scale=0.25]{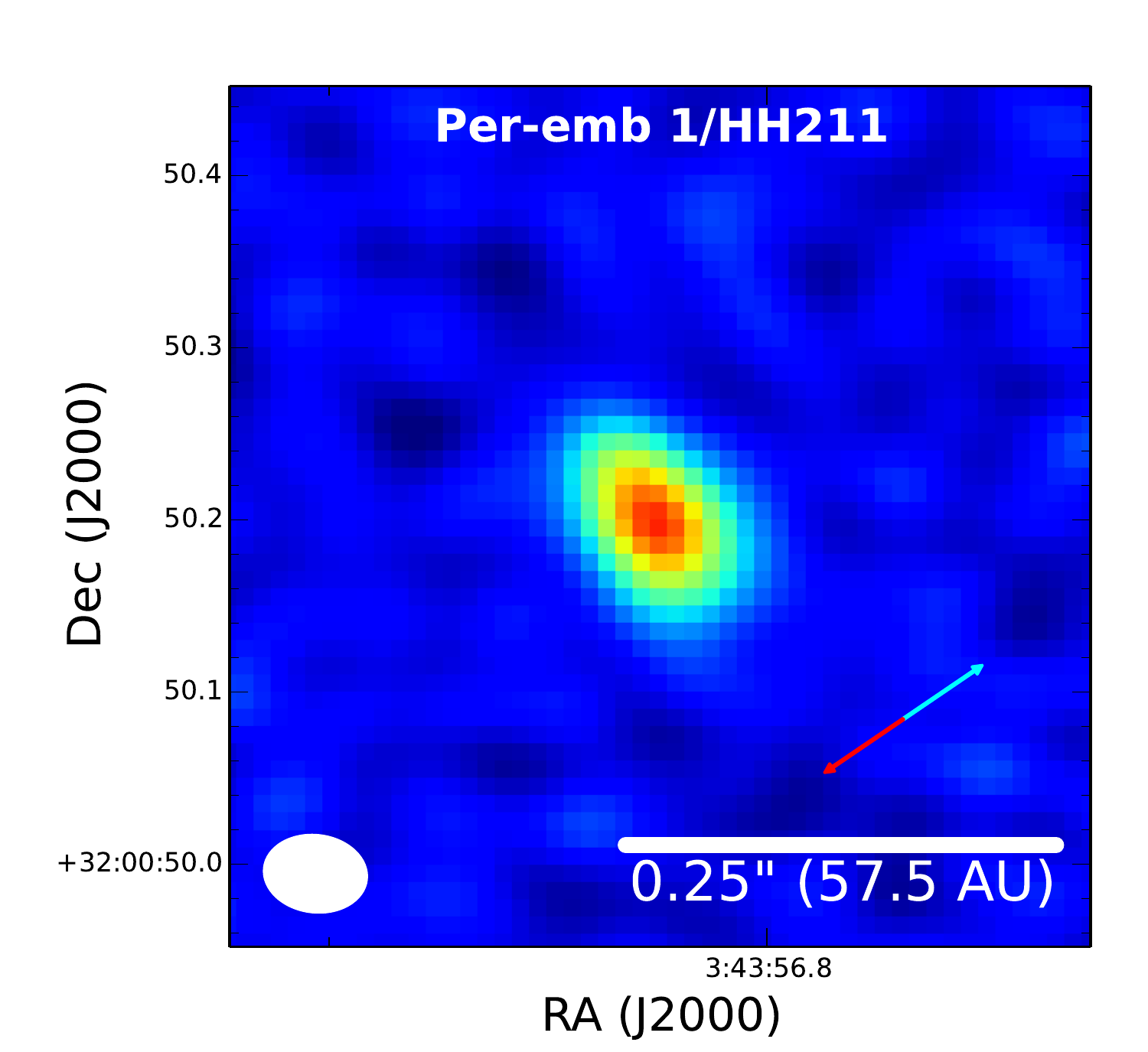}
\includegraphics[scale=0.25]{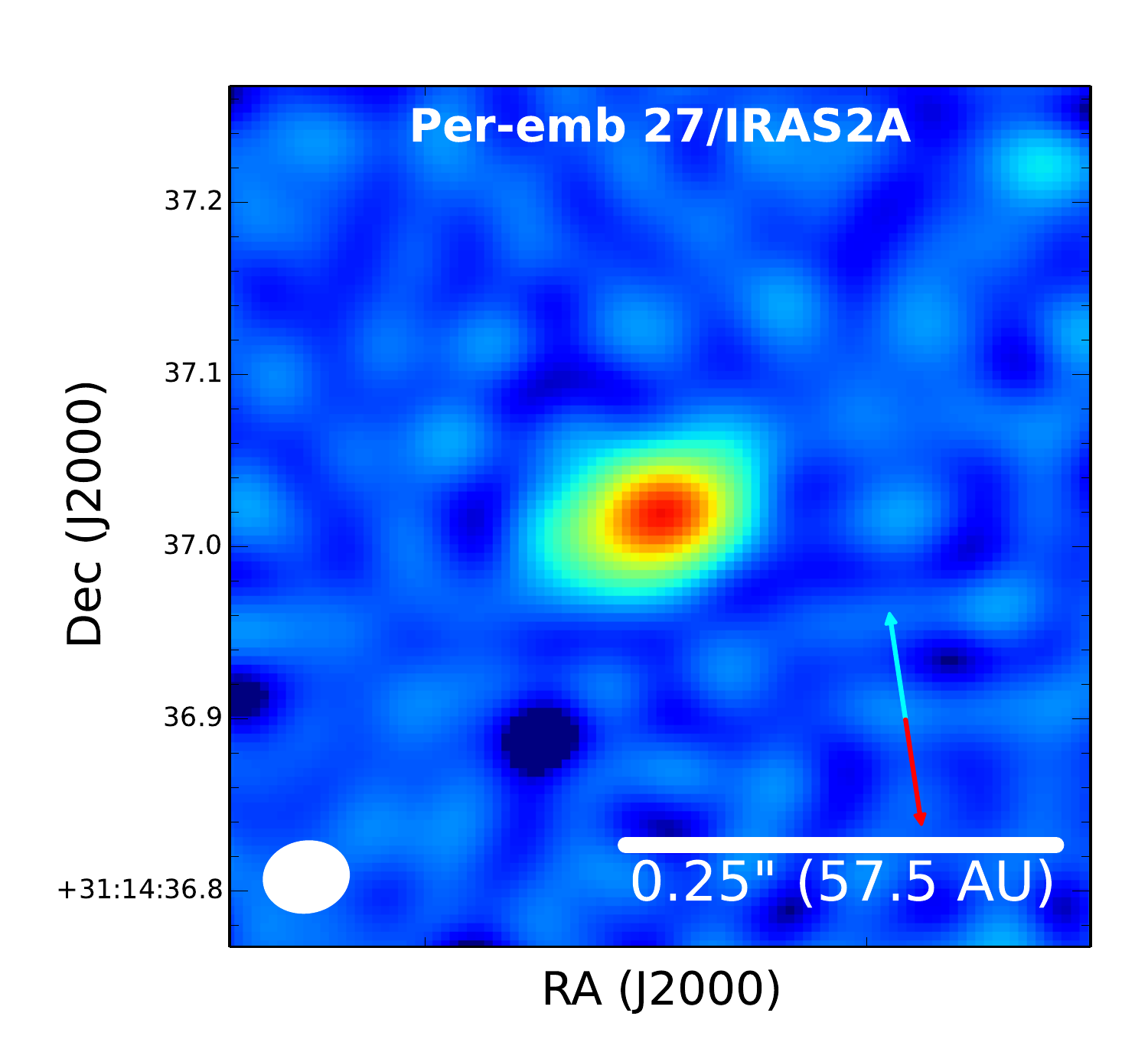}
\includegraphics[scale=0.25]{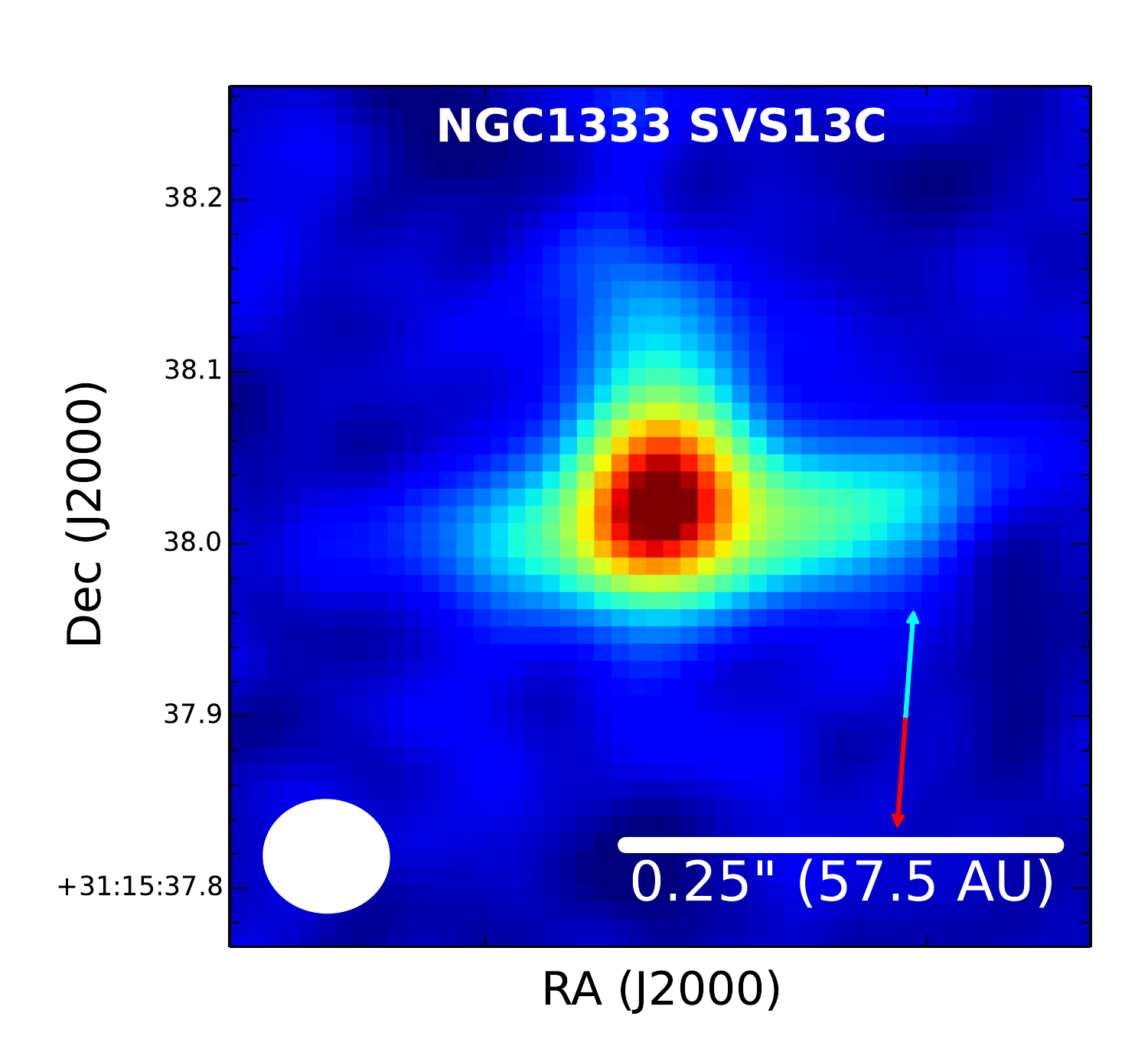}
\vspace{-5mm}
\end{center}

\caption{VLA 8~mm images of selected disk candidates. The
 continuum emission from these sources is compact, but their extensions
are perpendicular to the direction of the outflow, as expected for
disks. The outflow directions are drawn in the lower right.}
\end{figure}

\begin{figure}
\begin{center}
\includegraphics[scale=0.25]{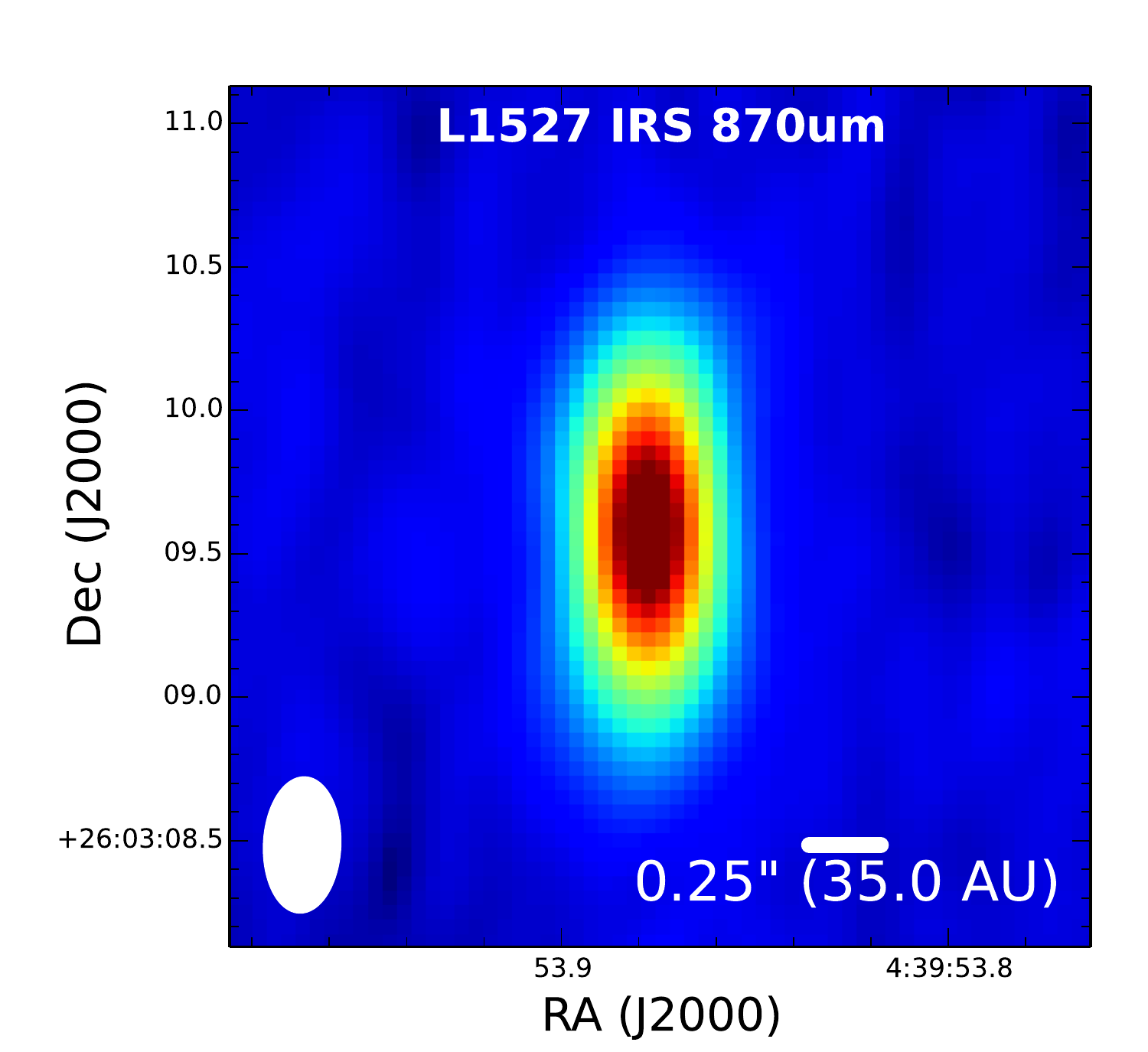}
\includegraphics[scale=0.25]{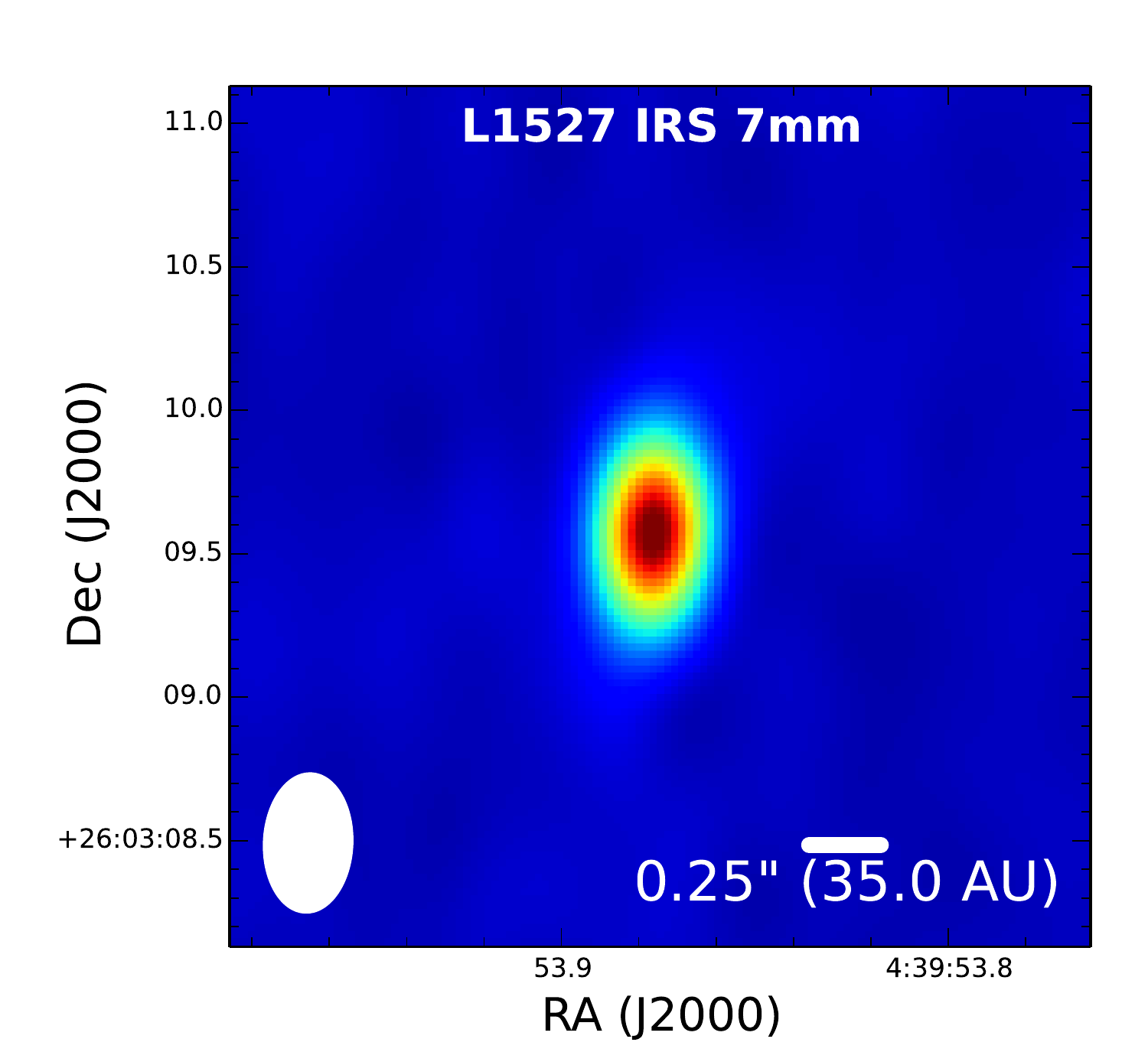}
\vspace{-5mm}
\end{center}
\caption{Continuum images of L1527 from ALMA (left) at 870 \micron\ and the VLA (right) at 7 mm from
Tobin et al. (in prep.), convolved to the larger restoring beam in the ALMA image. The deconvolved sizes 
are 0\farcs62 and 0\farcs26 at 870 \micron\ and 7 mm, respectively.}
\end{figure}

\section{Multiple Systems}

\begin{figure}
\begin{center}
\includegraphics[scale=0.25]{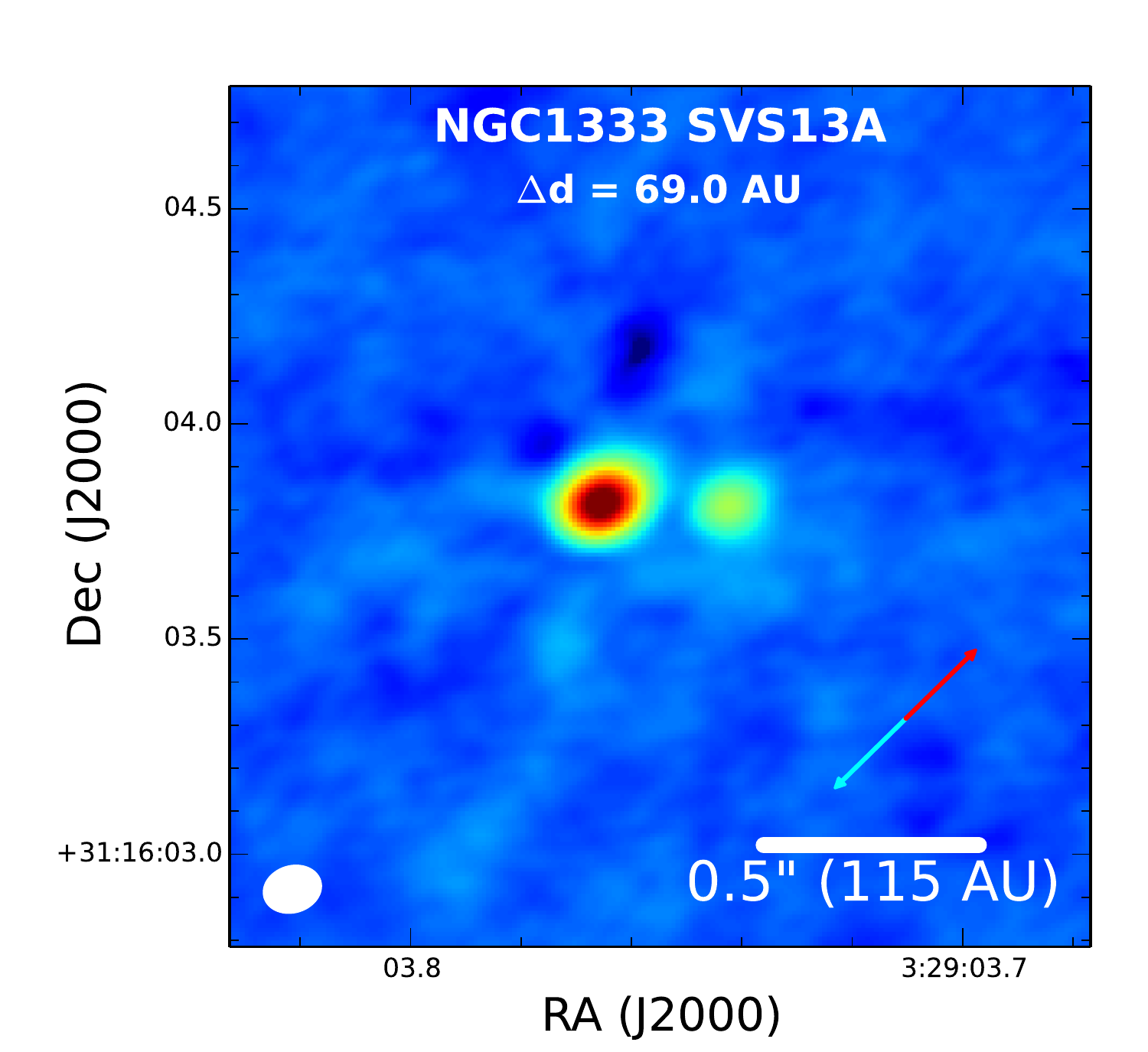}
\includegraphics[scale=0.25]{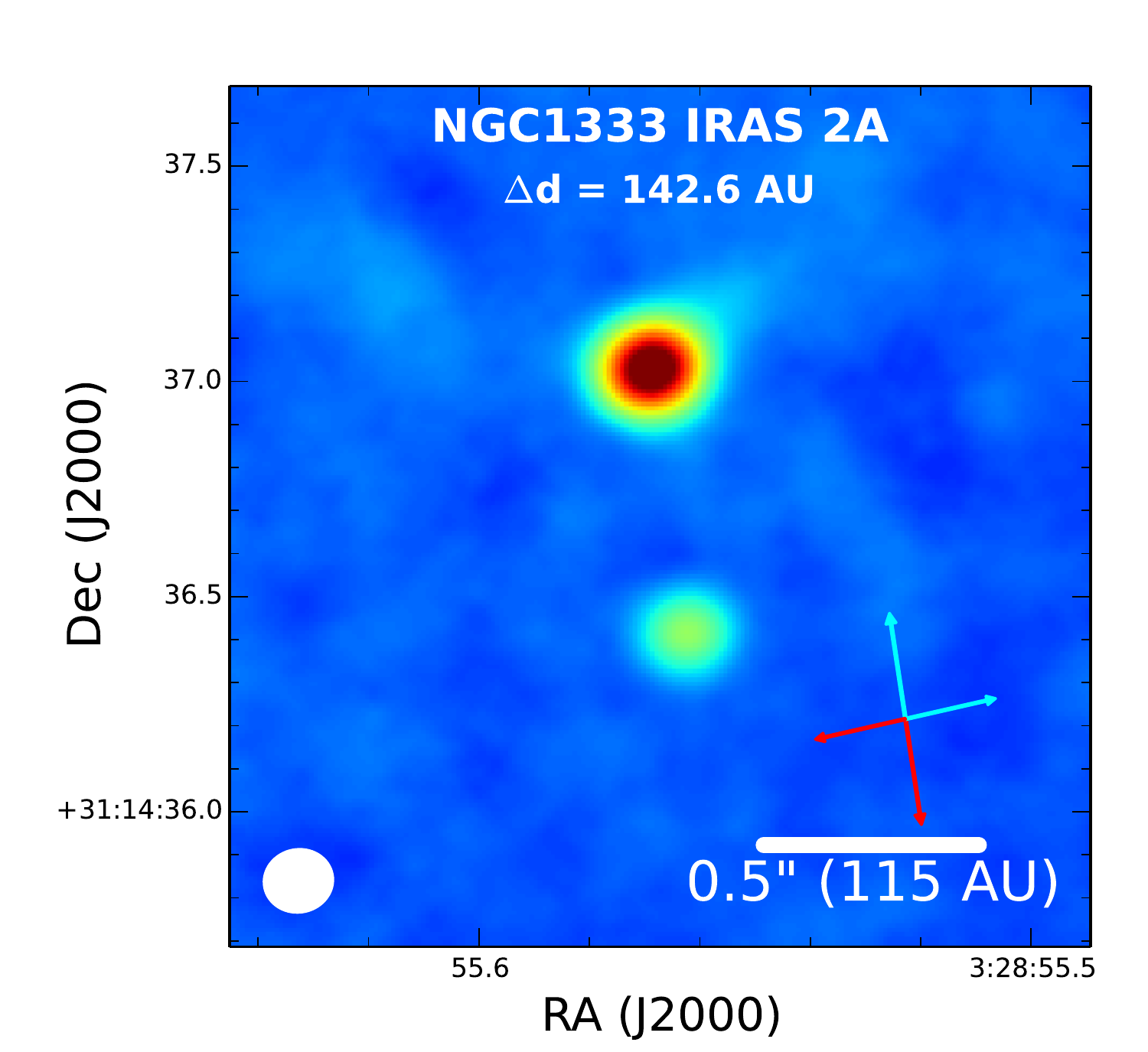}
\includegraphics[scale=0.25]{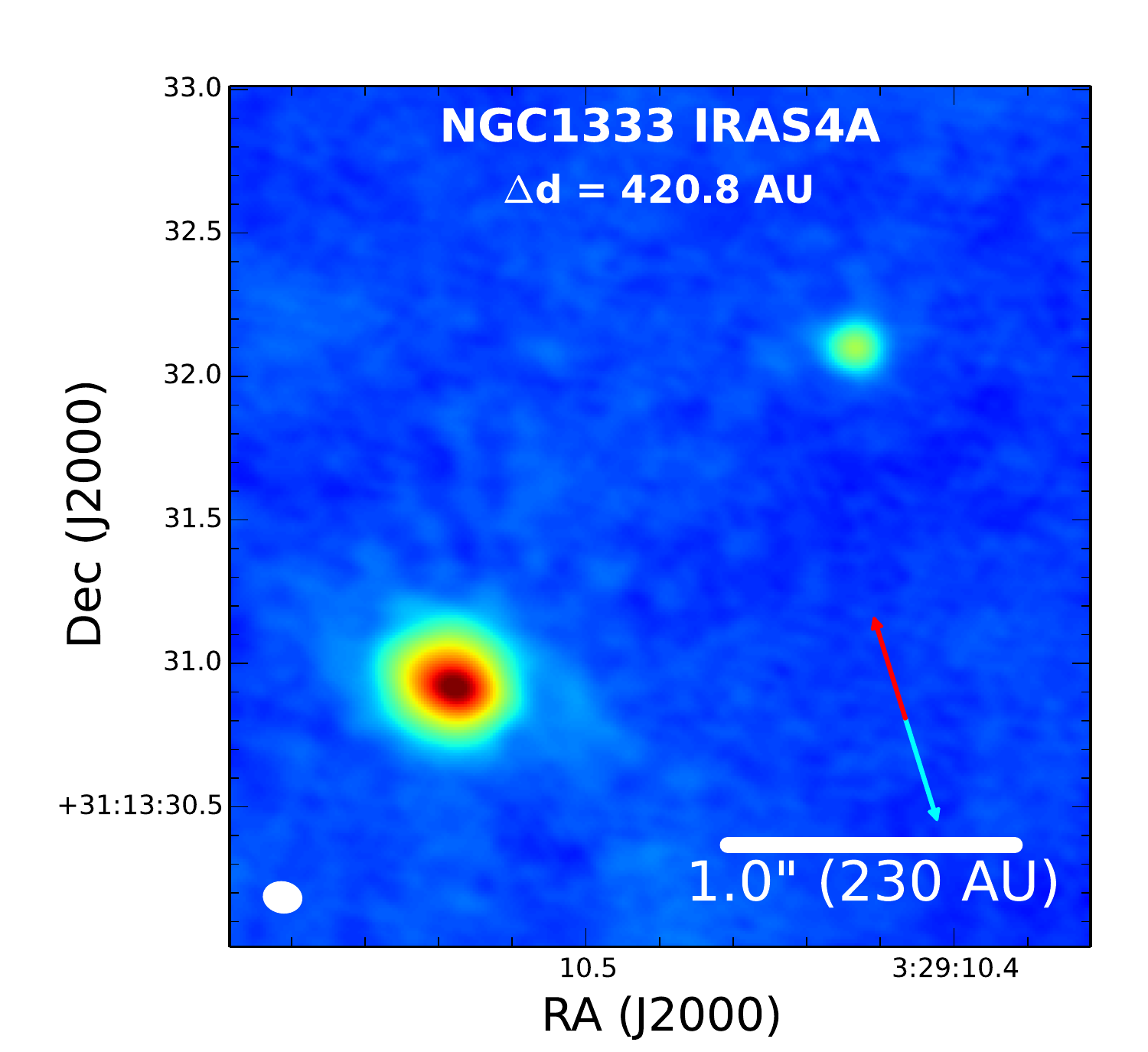}
\vspace{-5mm}
\end{center}
\caption{VLA 9~mm images of select multiple systems from the Perseus survey. The
outflow directions from the source(s) are plotted in the lower right corner.}
\end{figure}

With the ability to resolve multiples
down to separations of 15 AU, the VANDAM survey represents a 
significant evolution in our knowledge of protostellar 
multiplicity. We show examples of a few multiples systems from the VANDAM survey in Figure 3.
The companion of IRAS 2A, with a 142 AU separation, was first discovered in the VANDAM survey and is the apparent driving source
of the east-west outflow from the system (Tobin et al. 2015a). The SVS13A multiple
was first detected by Anglada et al. (2000) and IRAS 4A by Looney et al. (2000). Note that the brighter
source in IRAS 4A (4A1 or SE) is well resolved at 8 mm, possibly tracing a massive inner envelope or disk. 
From the 8 mm data, we have also detected linear polarization of the dust emission toward IRAS 4A1 (Cox et al. 2015),
tracing an apparent toroidal magnetic field in this system.

The full multiplicity results of the VANDAM survey are shown in Tobin et al. (submitted), where we observe a clear bi-modal
distribution of companion separations; one peak is at 75 AU and another is at $\sim$3000 AU. The 75 AU separation peak had
not been previously detected due to limited resolution, and we suggest that the 75 AU peak may result from disk fragmentation.
The 3000 AU separation peak may result from turbulent or Jeans fragmentation.

\section{Summary}
The key to the success of the VANDAM survey was in its unbiased nature and the superb 
sensitivity of the upgraded VLA, obtaining
as complete a characterization of protostellar disks and multiple systems as possible.

%%-----------------------------
%%      your bibliography
%%-----------------------------

\end{document}